\documentclass[fleqn,twoside]{article}
\usepackage{espcrc2,psfig}

\newlength{\figwidth}
\title{Study of spatial meson correlators at finite temperature 
in quenched anisotropic lattice QCD%
\thanks{Poster presented by K.~Nomura.}}

\author{
K.~Nomura\address{%
    Department of Physics, Hiroshima University, 
    Higashi-Hiroshima 739-8526, Japan \vspace{-0.2cm}},
O.~Miyamura$^{\rm a}$,
T.~Umeda\address{%
    Center for Computational Physics, University of Tsukuba,
    Tsukuba 305-8577, Japan \vspace{-0.2cm}},
and
H.~Matsufuru\address{%
    Yukawa Institute for Theoretical Physics, Kyoto University,
    Kyoto 606-8502, Japan} }

\begin{document}

\begin{abstract}
We analyze the meson correlator in the spatial direction at finite
temperature.
To achieve fine resolution in the spatial direction, we use an
anisotropic lattice
with the standard Wilson plaquette gauge action
and the $O(a)$ improved Wilson quark action.
Below and above $T_c$, properties of correlators are
investigated by two methods:
fits with ansatz for the spectral function,
and direct reconstruction of the spectral function using the
maximum entropy method.

\end{abstract}
\maketitle

\section{Introduction}



The extraction of the spectral functions from the correlators measured
in lattice QCD simulation is of great importance to understand the
QCD phase transition at finite temperature, both in theoretical
and phenomenological points of view.
Our goal is to investigate the both of spatial and temporal 
structure of hadron correlators near the deconfining transition
\cite{TARO01}.
Particularly for the latter, one needs to develop techniques
to extract detailed and reliable information from the correlators.
On the other hand, the correlators in the spatial direction are
rather easy to measure precisely, and therefore are expected
to serve as a good test ground for the method.
In this report, we focus on the spatial meson correlators
at finite temperature, and develop procedures to analyze them.
These analyses will give us insights also for the static screening
structure of QCD at $T>0$, which still remains not well understood
\cite{screening}.



We introduce a spectral function for the correlator propagating
in the spatial direction in analogous to the ordinary spectral
function obtained from the correlator propagating in the temporal
direction. In order to reconstruct this spectral function
from the lattice data, we apply two kinds of analysis procedures:
(a). Fit with ansatz on the shape of spectral function.
At present stage, we assume that the spectral function is
represented by a sum of several strong peaks, {\it i.e.}
delta functions, and fit the data to this form to obtain the screening
masses corresponding to the peaks.
We adopt single, two, and three pole(s) fittings.
(b). Reconstruction of spectral function from the correlator
by the maximum entropy method (MEM) \cite{NAH99}.
Comparing the results of these two methods, we discuss their
efficiency and reliability.

For these analysis, it is better to have
as many points of correlators as possible.
We use the anisotropic lattice which has small lattice spacing in
$z$-direction to measure the spatial hadron correlators
in high resolution.
Although one needs additional effort for the calibration of anisotropy 
parameters, once it has been done, the systematic uncertainties due to
the anisotropy can be controlled within reasonable accuracy
\cite{Aniso01b}.
In the following, we show only the results for the pseudoscalar
meson correlators, while the procedures are applicable for
other channel in the same manner.



\section{Numerical Simulations and Results}

Our numerical simulations are carried out on quenched lattices of
sizes $12^2\times 96\times N_t$ where $N_t=$12, 5, 3 and 2
which roughly correspond to the temperatures $\simeq 0$, $0.8T_c$,
$1.3T_c$ and $2T_c$, respectively.
The gauge configurations are generated with the Wilson plaquette
action at $\beta = 5.75$ with the anisotropy $\xi \!=\! a/a_z \!=\! 4$,
where $a_z$ and $a$ are the lattice spacings in z-direction and 
other three directions \cite{Aniso01b}.
With these parameters, the lattice cutoff set by the string tension
is $a^{-1}_{z} \simeq 4$ GeV.
Configuration numbers are 400 at $T\simeq 0$ and 510 at other $T$'s.
The quark action is the $O(a)$ improved Wilson action \cite{Aniso01b},
at three values of hopping parameter, $\kappa=$0.124, 0.122 and 0.120,
which cover the range of quark mass 
$m_q =(0.5 \sim 1.5)m_s$.
The bare quark anisotropy is set as $\gamma_F=$, which is
determined using the meson dispersion relation from the lattice
Klein-Gordon type action.

In the following, we show the results for the pseudoscalar meson
channel at $\kappa=0.120$ ($m_q\simeq 1.5 m_s$).
The fit is performed with the fitting region
$[L_{min}, L_{max}]$, where $L_{max}$ is fixed to $48$ and
$L_{min}$ is varied.
The screening masses as the results of single, double and triple
pole fits are displayed in the figures for each value
of $L_{min}$.
Existence of a region where the masses are stable  indicates that 
the assumed form of fit is a good representation of the spectral function.
On the other hand, absence of such a region signals that the spectral
function is not approximated by such a form.
In both cases,  we can extract the spectral function
reconstructed by MEM as an alternative procedure.
In MEM analysis we use the default model function 
$m_0(\omega)=16 \omega^2$, frequency region $0 \leq \omega \leq 2.0$
and the fitting region of correlators $1 \leq t \leq 40$.

\begin{figure}[tb]
\vspace{0.3cm}
\centerline{\psfig{figure=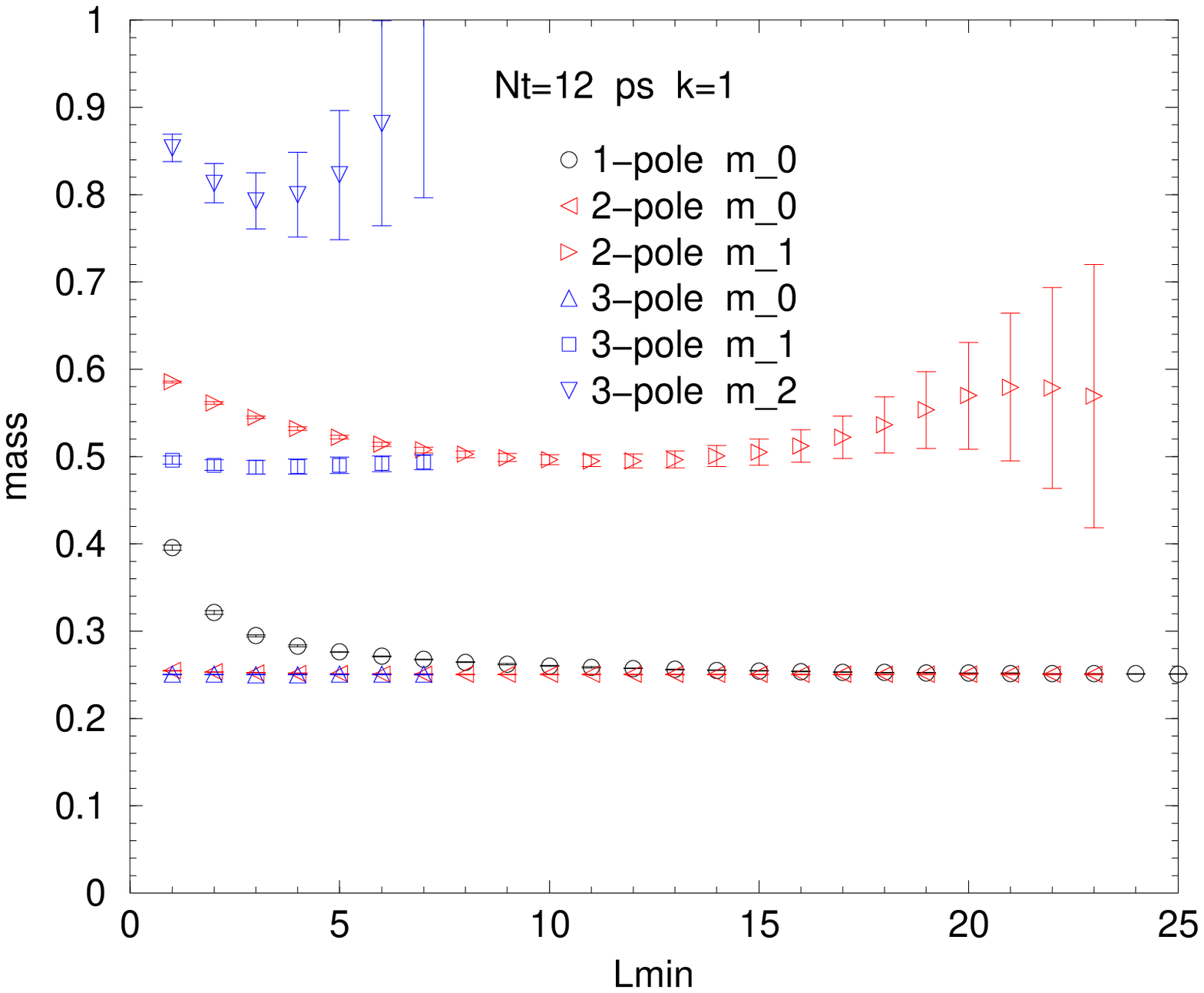,width=\figwidth}}
\centerline{\psfig{figure=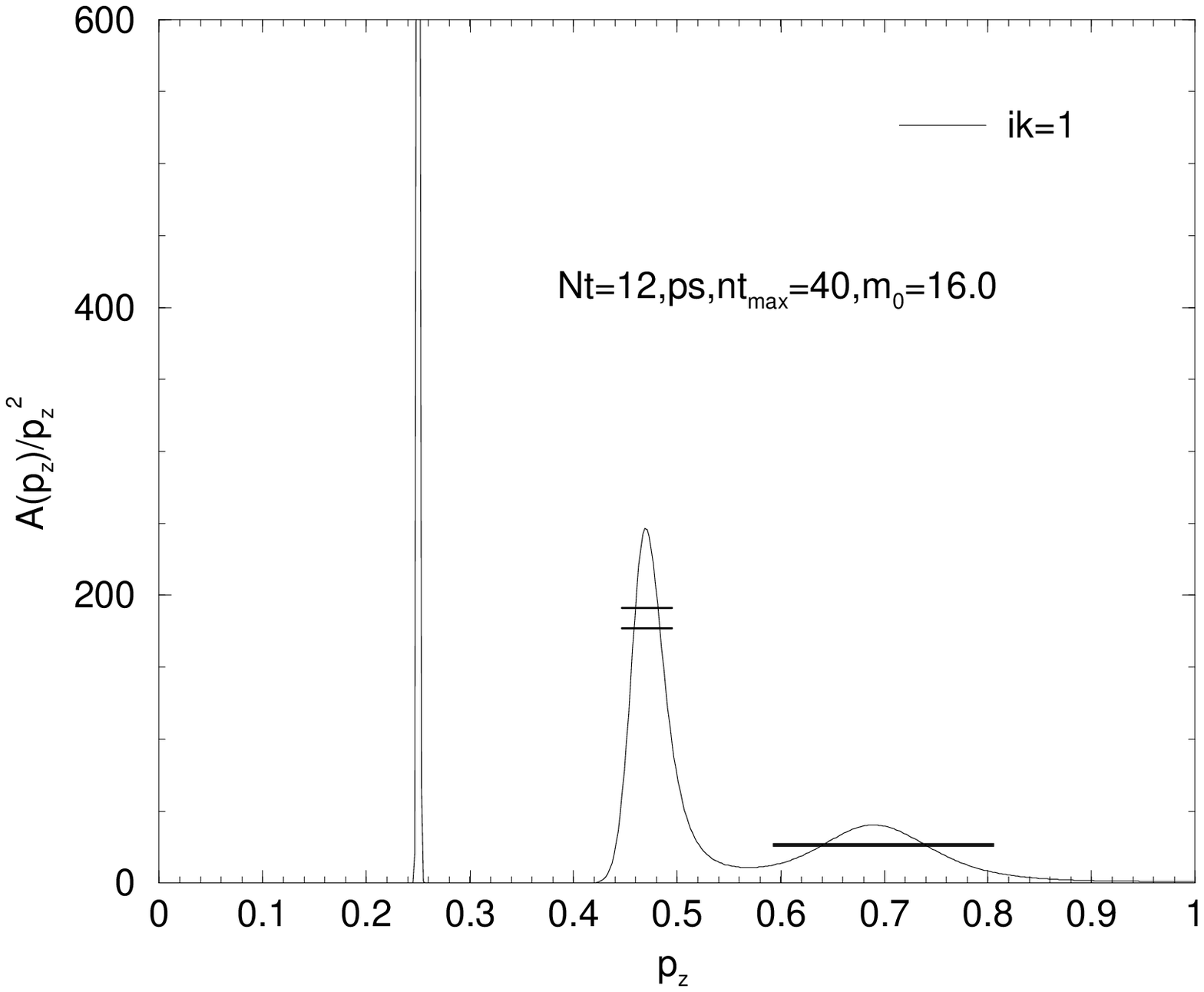,width=\figwidth}}
\vspace{-1.0cm}
\caption{Results at $T=0$.
The upper panel is the result of fits.
The lower panel shows the spectral function obtained using MEM.}
\label{fig:zero_temp}
\vspace{-0.4cm}
\end{figure}


Let us start with the zero temperature.
The upper panel of Figure~\ref{fig:zero_temp} shows the results
fitting analysis.
For each fit (\textit{i.e.} number of poles), there is a region
where the masses do not change within errors.
For the ground and first excited states,  three fits give
consistent masses.
This shows that the assumed form of spectral function
(the sum of poles) well represent the correlator, in accord
with the physical expectation.
The spectral function reconstructed by MEM also exhibits consistent
position of peaks, as shown in the lower panel of
Fig.~\ref{fig:zero_temp}.
Since the masses from the fits shows clear plateaus, and observing
the errors associated to the regions of spectral function,
the observed widths of peaks may not be practical ones 
but artifacts through the statistical and systematic errors MEM.

\begin{figure}[tb]
\vspace{0.3cm}
\centerline{\psfig{figure=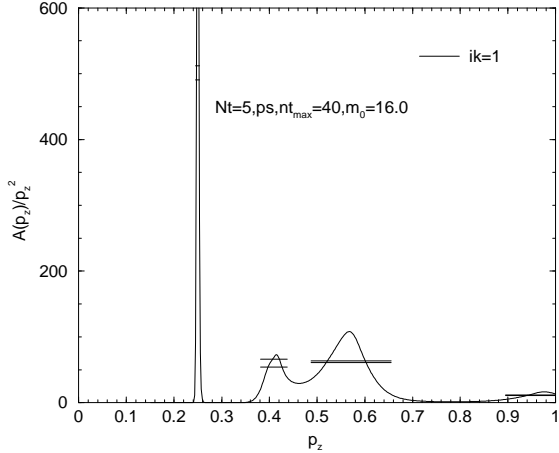,width=\figwidth}}
\vspace{-1.0cm}
\caption{Result for the spectral function at $T=0.8 T_c$.}
\label{fig:below_Tc}
\vspace{-0.4cm}
\end{figure}


Now we apply the same procedures to the correlators at finite
temperature.
The result of fit method at $0.8 T_c$ is almost the same as at $T=0$.
Therefore there seems no significant change in the spectral functions
at $T=0.8T_c$ from that at $T=0$.
On the other hand, as shown in Figure~\ref{fig:below_Tc},
the result of MEM shows a deference:
there is two peaks at the position of the first excited state obtained
by the fits.
This is unphysical structure can be understood by the
failure of MEM in reconstructing the spectral function, probably
due to lack of statistics.
In order to detect the structure of spectral function in high
frequency region from MEM, one would probably need 
larger statistics that the present one which of $O(500)$.

\begin{figure}[tb]
\vspace{0.3cm}
\centerline{\psfig{figure=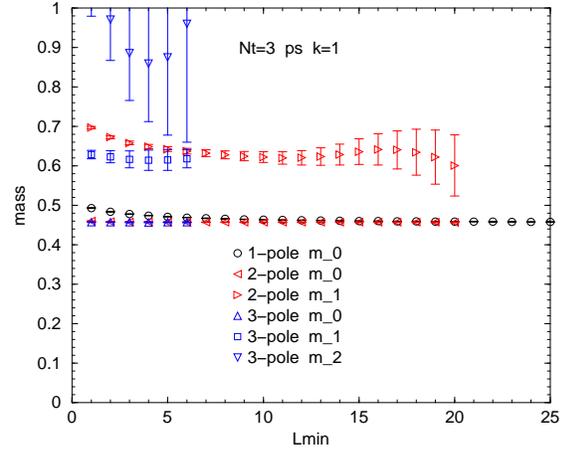,width=\figwidth}}
\centerline{\psfig{figure=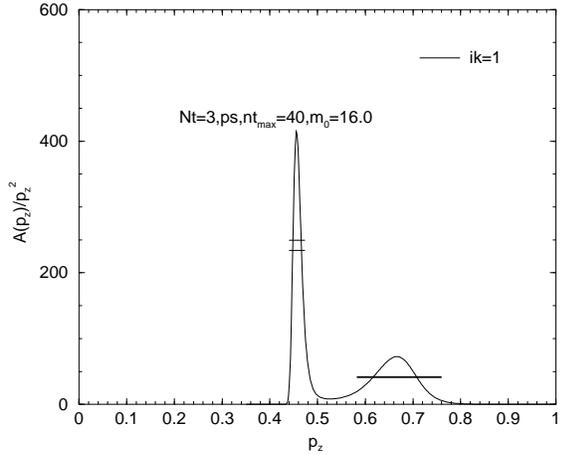,width=\figwidth}}
\vspace{-1.0cm}
\caption{Results at $T=1.3 T_c$.}
\label{fig:above_Tc}
\vspace{-0.4cm}
\end{figure}


Figure~\ref{fig:above_Tc} shows the results at $1.3 T_c$.
We again find the regions of constant masses
for each fit and a consistency in the masses of the ground
and first excited states.
Result of spectral function also have peaks at consistent positions.
Thus we find that the sum of poles are a good approximation of
the spectral function.
The ground state mass is slightly less than the twice lowest
Matsubara frequency.
For the quantitative analysis, however, present lattice is so
coarse that the number of temporal degrees of freedom is just three.
We also observe similar result at $T=2 T_c$.

We conclude that these two procedures are promising as 
useful ways to extract the structure of the spectral function,
although one need careful treatment particularly for MEM.
Fits assuming various forms for the spectral function serve
complementary way of analysis, and supply the reliability for
the result of MEM.
Such a study is particularly important for the analysis of
temporal correlators, since the degrees of freedom is restricted
even with anisotropic lattices and the data are inevitably
in the high frequency region.
In this case, since the spectral function have physical width,
one also need to try other fitting forms, such as
the peaks with widths.
To obtain more quantitative result on the screening mass and 
the spectral function for the spatial correlators, we should 
use smaller lattice spacing as well as higher statistics.

\end{document}